\documentclass[showpacs,twocolumn,floatfix]{revtex4}
\usepackage{graphicx}
\usepackage{amsmath}
\usepackage{amssymb}
\begin{document}
\title{Entangling ability of a beam splitter in the presence of temporal which-path information}
\author{J. L. van Velsen}
\affiliation{Instituut-Lorentz, Universiteit Leiden, P.O. Box 9506, 2300 RA Leiden, The Netherlands}
\date{March 2005}
\begin{abstract}
We calculate the amount of polarization-entanglement induced by two-photon interference at a lossless beam splitter. 
Entanglement and its witness are quantified respectively by concurrence and the Bell-CHSH parameter.
In the presence of a Mandel dip, the interplay of two kinds of which-path information --- temporal and polarization --- gives rise to the
existence of entangled polarization-states that cannot violate the Bell-CHSH inequality.  
\end{abstract}
\maketitle

\section{Introduction}

Entanglement, the nonclassical correlations between spatially separated particles, is typically a signature of interactions in the past or
emergence from a common source. However, it can also arise as the interference of identical particles \cite{Yur92}.
By postselecting experimental data based on the  ``click'' of detectors \cite{Shi88,Ou88}, photons scattered at a beam splitter have violated a Bell 
inequality, even if they originated from independent sources \cite{Pit03,Fat04}. In reverse, triggered by an interferometric 
Bell-state measurement, entanglement has been swapped \cite{Zuk93} to initially uncorrelated photons of different Bell pairs \cite{Pan98a,Pan98b,Jen02}. 
The observation of these nonclassical interference effects is an important step on the road towards an optical approach of quantum information 
processing \cite{Kni01,Fra02}. 

Being furnished by interference, the ability of a beam splitter to entangle the polarizations of two independent photons depends on their 
indistinguishability \cite{Fey69}. One of the incident photons is horizontally polarized in state $|{\rm H};\psi\rangle$, the other 
vertically polarized in $|{\rm V};\phi\rangle$. The photons are partially distinguishable by their temporal degrees of freedom captured 
in the kets $|\psi\rangle$ and $|\phi\rangle$. Besides temporal which-path information inherited from incident photons, a scattered
two-photon state possibly holds polarization which-path information. We make no assumptions about the scattering amplitudes 
connecting polarizations at the beam splitter, except that they constitute a unitary scattering matrix.   
Translated to a polarization-conserving beam splitter, this corresponds to incident photons in states $|\sigma;\psi\rangle$ and 
$|\sigma';\phi\rangle$ where $\sigma$, $\sigma'$ are arbitrary superpositions of ${\rm H}$, ${\rm V}$.
Our analysis generalizes existing work on a polarization-conserving beam splitter where $\sigma={\rm H}$ and $\sigma'={\rm V}$ \cite{Bos02,Fat04}.

The polarization-state $\rho$ of a scattered photon pair is established from the scattering amplitudes of the beam splitter, the shape and timing of 
photonic wavepackets ($|\psi\rangle$, $|\phi\rangle$) and the time-window of coincidence detection. If not erased by  
ultra-coincidence detection, an amount of temporal distinguishability of $(1-|\langle \psi|\phi \rangle|^{2})$ pertains corresponding to a mixed state 
$\rho$. We calculate both its concurrence and the Bell-CHSH parameter.
The ability of the latter to witness entanglement can disappear in the presence of a Mandel dip. 
In terms of a polarization-conserving beam splitter, this corresponds to a deviation of $\sigma$, $\sigma'$ from
$\sigma={\rm H}$ and $\sigma'={\rm V}$. 

\section{Formulation of the problem}

\begin{figure}
\includegraphics[width=8cm]{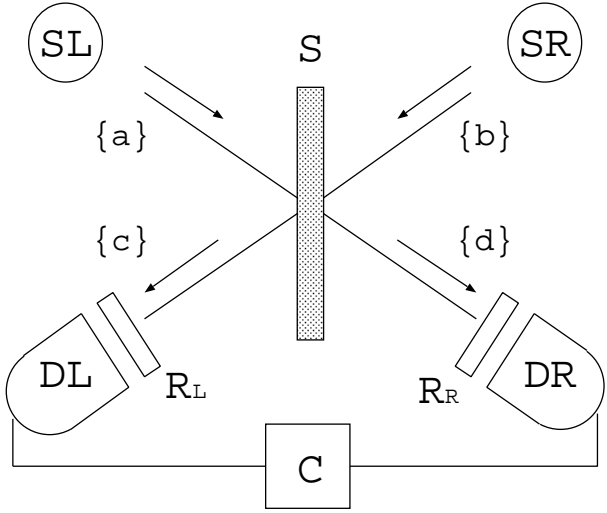}
\caption{
Schematic drawing of generation and detection of polarization-entanglement at a beam splitter. 
The independent sources SL and SR each create a photon in modes $\{a\}$ and $\{b\}$ cf. Eq. (\ref{Psiin}). 
The beam splitter with unitary $4 \times 4$ scattering matrix $S$ couples the polarization of incoming modes to the 
polarization of outgoing modes $\{c\}$ and $\{d\}$. Polarizations are mixed by $R_{\rm L},R_{\rm R}$ and absorbed by photodetectors
DL,DR. A coincidence circuit C registers simultaneous detection of photons.  
\label{beamsplitter}
}
\end{figure}

In a second-quantized notation, the incident two-photon state $|{\rm H};\psi\rangle_{\rm L}|{\rm V};\phi\rangle_{\rm R}$ takes the form
\begin{equation}
|{\rm \Psi}_{\rm in}\rangle=
{\rm \Psi}_{\rm H,L}^{\dagger}{\rm \Phi}_{\rm V,R}^{\dagger}|0\rangle, 
\label{Psiin}
\end{equation}
with field creation operators given by (see Fig. \ref{beamsplitter})
\begin{equation}
{\rm \Psi}_{\rm H,L}^{\dagger}=\int d\omega\, a_{\rm H}^{\dagger}(\omega)\psi^{*}(\omega), \quad
{\rm \Phi}_{\rm V,R}^{\dagger}=\int d\omega\, b_{\rm V}^{\dagger}(\omega)\phi^{*}(\omega).
\end{equation} 
(The subscripts R,L indicate the two sides of the beam splitter.)
The operators $a_{i}(\omega)$ with $i={\rm H},{\rm V}$ satisfy commutation rules
\begin{equation}
[a_{i}(\omega),a_{j}(\omega')]=0, \quad [a_{i}(\omega),a_{j}^{\dagger}(\omega')]=\delta_{ij}\delta(\omega-\omega').
\end{equation}
The same commutation rules hold for the operators $b_{i}(\omega)$, with commutation among  
$a$ and $b$. 

The outgoing operators $c_{i}(\omega)$, $d_{i}(\omega)$ are related to the incoming ones
$a_{i}(\omega)$, $b_{i}(\omega)$ by a $4 \times 4$ unitary scattering matrix $S$, decomposed in
$2 \times 2$ reflection and transmission matrices $r$,$t$,$t'$,$r'$:
\begin{equation}
\left(\begin{array}{c} c(\omega) \\ d(\omega) \end{array} \right) = 
\left( \begin{array}{ll}
r & t' \\
t & r' \end{array} \right)
\left(\begin{array}{c} a(\omega) \\ b(\omega) \end{array} \right), \quad 
a(\omega) \equiv \left(\begin{array}{c} a_{\rm H}(\omega) \\ a_{\rm V}(\omega) \end{array} \right),  
\label{inout}
\end{equation}
and vectors $b(\omega)$, $c(\omega)$, $d(\omega)$ defined similarly. The scattering amplitudes are   
frequency-independent. 
\begin{widetext}
The outgoing state $|{\rm \Psi}_{\rm out}\rangle$ can be conveniently written in a matrix notation 
\begin{equation}
|{\rm \Psi}_{\rm out}\rangle=
\int d\omega \int d\omega' \, \psi^{*}(\omega)\phi^{*}(\omega')
\left(\begin{array}{c} c^{\dagger}(\omega) \\ d^{\dagger}(\omega) \end{array} \right)^{\rm T}
\left( \begin{array}{ll}
r\sigma_{\rm in}t'^{\rm T} & r\sigma_{\rm in}r'^{\rm T} \\
t\sigma_{\rm in}t'^{\rm T} & t\sigma_{\rm in}r'^{\rm T}  \end{array} \right)
\left(\begin{array}{c} c^{\dagger}(\omega') \\ d^{\dagger}(\omega') \end{array} \right)|0\rangle.
\label{psimatout}
\end{equation}
\end{widetext}
Here we used the unitarity of $S$ and $\sigma_{\rm in}=(\sigma_{x}+i\sigma_{y})/2$, with $\sigma_{x}$ and $\sigma_{y}$ Pauli matrices,    
corresponds to the polarizations of the incoming photons cf. Eq. (\ref{Psiin}).
The matrix $\sigma_{\rm in}$ has rank 1 reflecting the fact that 
polarizations are not entangled prior to scattering. Since we make no assumptions about the scattering amplitudes (apart from the unitarity of $S$), 
the choice of $\sigma_{\rm in}$ is without loss of generality (see Appendix \ref{stostate}).      

The joint probability per unit $\mbox{(time)}^{2}$ of absorbing a photon with polarization $i$ at detector DL and a photon with 
polarization $j$ at detector DR at times $t$ and  $t'$ respectively is given by \cite{Gla63}
\begin{equation}
w_{ij}(t,t') \propto \langle {\rm \Psi}_{\rm out}|E_{i{\rm L}}^{(-)}(t)E_{j{\rm R}}^{(-)}(t')E_{j{\rm R}}^{(+)}(t')E_{i{\rm L}}^{(+)}(t)
|{\rm \Psi}_{\rm out}\rangle,
\end{equation}
where $E_{i{\rm L}}^{(+)}(t)$ and $E_{i{\rm R}}^{(+)}(t)$ are the positive frequency field operators of polarization $i$ at detectors DL and DR.
The probability $C_{ij}(t)$ of a coincidence event within time-windows $\tau$ around $t$ is given by
\begin{equation}
C_{ij}(t)=\int_{t-\frac{\tau}{2}}^{t+\frac{\tau}{2}} dt'\int_{t-\frac{\tau}{2}}^{t+\frac{\tau}{2}} dt''w_{ij}(t',t'').
\label{Corstart}
\end{equation}  
Experimentally, the time-window $\tau$ has typically a lower bound determined by the random rise time of an avalanche of charge
carriers in response to a photon absorption event.

The polarization-entanglement is detected by violation of the Bell-CHSH inequality \cite{Cla69}. This requires two local polarization mixers
$R_{\rm L}$ and $R_{\rm R}$. The Bell-CHSH parameter ${\cal E}$ is 
\begin{equation}
{\cal E}=|E(R_{\rm L},R_{\rm R})+E(R'_{\rm L},R_{\rm R})+E(R_{\rm L},R'_{\rm R})-E(R'_{\rm L},R'_{\rm R})|,
\label{Bellpar}
\end{equation}
where $E(R_{\rm L},R_{\rm R})$ is related to the correlators $C_{ij}(R_{\rm L},R_{\rm R})$ by
\begin{equation}
E=\frac{C_{\rm HH}+C_{\rm VV}-C_{\rm HV}-C_{\rm VH}}{C_{\rm HH}+C_{\rm VV}+C_{\rm HV}+C_{\rm VH}}.
\label{Edef}
\end{equation}
Substituting the correlators of Eq. (\ref{Corstart}) into Eq. (\ref{Edef}), we see that 
\begin{equation}
E(R_{\rm L},R_{\rm R})={\rm Tr}\, \rho \, (R_{\rm L}^{\dagger}\sigma_{z}R_{\rm L}) \otimes (R_{\rm R}^{\dagger}\sigma_{z}R_{\rm R}), 
\label{Erho}
\end{equation}
where $\sigma_{z}$ is a Pauli matrix and $\rho$ a $4 \times 4$ polarization density matrix with elements
\begin{widetext}
\begin{equation}
\rho_{ij,mn} = \frac{1}{\mathcal{N}} \left( (1 + |\alpha|^{2})(\gamma_{1})_{ij}(\gamma_{1})^{*}_{mn}+(1 - |\alpha|^{2})
(\gamma_{2})_{ij}(\gamma_{2})^{*}_{mn} \right). 
\end{equation}
The parameter $\alpha$ is given by
\begin{equation}
\alpha = \frac{\int_{t-\frac{\tau}{2}}^{t+\frac{\tau}{2}} dt' \int d\omega \int d\omega' 
\phi(\omega)\psi^{*}(\omega')e^{i(\omega-\omega')t'}}
{\sqrt{\left(\int_{t-\frac{\tau}{2}}^{t+\frac{\tau}{2}} dt' \int d\omega \int d\omega' \phi(\omega)\phi^{*}(\omega')
e^{i(\omega-\omega')t'}\right)
\left(\int_{t-\frac{\tau}{2}}^{t+\frac{\tau}{2}} dt' \int d\omega \int d\omega' \psi(\omega)\psi^{*}(\omega')e^{i(\omega-\omega')t'}\right)}}
\end{equation}
\end{widetext}
and $\gamma_{1}$,$\gamma_{2}$ are $2 \times 2$ matrices related to the scattering amplitudes by
\begin{equation}
\gamma_{1} = r\sigma_{\rm in}r'^{\rm T}+t'\sigma_{\rm in}^{\rm T}t^{\rm T}, \quad 
\gamma_{2} = r\sigma_{\rm in}r'^{\rm T}-t'\sigma_{\rm in}^{\rm T}t^{\rm T}.
\label{gammas}
\end{equation} 
The normalization factor $\mathcal{N}$ takes the form
\begin{equation}
\mathcal{N}=(1 + |\alpha|^{2}){\rm Tr}\,\gamma_{1}^{\dagger}\gamma_{1}^{\vphantom{\dagger}}+
(1 - |\alpha|^{2}){\rm Tr}\,\gamma_{2}^{\dagger}\gamma_{2}^{\vphantom{\dagger}}.
\label{norm}
\end{equation}

The parameter $1-|\alpha|^{2} \in (0,1)$ represents the amount of temporal which-path information.
Generally, the time-window $\tau$ is much larger than the coherence times or temporal difference of the wavepackets. We may then take the
limit $\tau \rightarrow \infty$ and $\alpha$ reduces to the overlap of wavepackets
\begin{equation}
\alpha=\int d\omega \phi(\omega)\psi^{*}(\omega).
\label{overlap}
\end{equation} 
In the opposite limit of ultra-coincidence detection where $\tau \rightarrow 0$, 
temporal which-path information is completely erased corresponding to $|\alpha|^{2}=1$.

\section{Entanglement of formation}
\label{EOF}

The entanglement of formation of the mixed state $\rho$ is quantified by the concurrence $\mathcal{C}$ \cite{Woo98} given by
\begin{equation}
\mathcal{C}={\rm max}\left(0,\sqrt{\lambda_{1}}-\sqrt{\lambda_{2}}-\sqrt{\lambda_{3}}-\sqrt{\lambda_{4}}\right).
\end{equation}
The $\lambda_{i}$'s are the eigenvalues of the matrix product $\rho\tilde{\rho}$, where 
$\tilde{\rho}=(\sigma_{y}\otimes\sigma_{y})\rho^{*}(\sigma_{y}\otimes\sigma_{y})$,
in the order $\lambda_{1} \ge \lambda_{2} \ge \lambda_{3} \ge \lambda_{4}$. The concurrence ranges from 0 (no entanglement) to
1 (maximal entanglement).
For simplicity of notation it is convenient to define $(\widehat{xy})_{ij,mn} \equiv x_{ij}y^{*}_{mn}$. The matrix $\tilde{\rho}$ can be 
written as
\begin{equation}
\tilde{\rho} = \frac{1}{\mathcal{N}} \left( (1 + |\alpha|^{2})\widehat{\tilde{\gamma}_{1}\tilde{\gamma}_{1}}+
(1 - |\alpha|^{2})\widehat{\tilde{\gamma}_{2}\tilde{\gamma}_{2}} \right),
\end{equation} 
with $\tilde{\gamma} \equiv \sigma_{y}\gamma^{*}\sigma_{y}$.
The product $\rho\tilde{\rho}$ takes the simple form
\begin{equation}
\rho\tilde{\rho}=\frac{{\rm Tr}\,\gamma_{1}^{\dagger}\tilde{\gamma}_{1}^{\vphantom{\dagger}}}{\mathcal{N}^{2}}
\left((1 + |\alpha|^{2})^{2}\widehat{\gamma_{1}\tilde{\gamma}_{1}}-
(1 - |\alpha|^{2})^{2}\widehat{\gamma_{2}\tilde{\gamma}_{2}} \right),
\end{equation}
where we have used the multiplication rule $\widehat{xy}\widehat{vw}=({\rm Tr}\,y^{\dagger}v) \widehat{xw}$ and 
\begin{equation}
{\rm Tr}\,\gamma_{1}^{\dagger}\tilde{\gamma}_{1}^{\vphantom{\dagger}}=-{\rm Tr}\,\gamma_{2}^{\dagger}
\tilde{\gamma}_{2}^{\vphantom{\dagger}}, \quad
{\rm Tr}\,\gamma_{1}^{\dagger}\tilde{\gamma}_{2}^{\vphantom{\dagger}}={\rm Tr}\,\gamma_{2}^{\dagger}
\tilde{\gamma}_{1}^{\vphantom{\dagger}}=0.
\label{traces1}
\end{equation}
The results for the tilde inner products of Eq. (\ref{traces1}) hold since the photons are not polarization-entangled
prior to scattering (${\rm Det}\,\sigma_{\rm in}=0$). 

\begin{widetext}
The non-Hermitian matrix $\rho\tilde{\rho}$ has eigenvalue-eigenvector decomposition
\begin{equation}
\rho\tilde{\rho}=\frac{|{\rm Tr}\,\gamma_{1}^{\dagger}\tilde{\gamma}_{1}^{\vphantom{\dagger}}|^{2}}{\mathcal{N}^{2}}
\left(\sum_{i=1,2} \widehat{\gamma_{i}s_{i}}\right) 
\left( (1 + |\alpha|^{2})^{2}\widehat{s_{1}s_{1}}+
(1 - |\alpha|^{2})^{2}\widehat{s_{2}s_{2}} \right)
\left(\sum_{i=1,2} \widehat{\gamma_{i}s_{i}}\right)^{-1},
\label{simtrans}
\end{equation}
\end{widetext}
where we have defined orthonormal states  
$s_{1}=(1/2)(\openone+\sigma_{z})$ and $s_{2}=(1/2)(\sigma_{x}+i\sigma_{y})$.
The pseudo-inverse is easily seen to be
\begin{equation}
\left(\sum_{i=1,2} \widehat{\gamma_{i}s_{i}}\right)^{-1}=
\frac{1}{({\rm Tr}\,\gamma_{1}^{\dagger}\tilde{\gamma}_{1}^{\vphantom{\dagger}})^{*}}
\left(\widehat{s_{1}\tilde{\gamma}_{1}}-\widehat{s_{2}\tilde{\gamma}_{2}}\right).
\end{equation} 
It follows that
\begin{equation}
\mathcal{C}=\frac{2|\alpha|^{2}|{\rm Tr}\,\gamma_{1}^{\dagger}\tilde{\gamma}_{1}^{\vphantom{\dagger}}|}{\mathcal{N}}.
\label{Ctemp}
\end{equation}
The trace that appears in the numerator of Eq. (\ref{Ctemp}) is given by
\begin{equation}
|{\rm Tr}\,\gamma_{1}^{\dagger}\tilde{\gamma}_{1}^{\vphantom{\dagger}}|  = 
2\sqrt{{\rm Det}\,X^{\dagger}X\,{\rm Det}(\openone-X^{\dagger}X)}, 
\label{trace2}
\end{equation}
where we have defined a ``hybrid'' $2 \times 2$ matrix $X$ as
\begin{equation}
X=\left(\begin{array}{cc} r_{\rm HH} & t'_{\rm HV} \\ r_{\rm VH} & t'_{\rm VV} \end{array}\right).
\end{equation}
The normalization factor $\mathcal{N}$ given by Eq. (\ref{norm}) can be expressed in terms of $X$ using
\begin{equation}
{\rm Tr}\,\gamma_{1}^{\dagger}\gamma_{1}^{\vphantom{\dagger}}  =  {\rm Tr}\,X^{\dagger}X-2\,{\rm Per}\,X^{\dagger}X, 
\label{trace3}
\end{equation}
\begin{equation}
{\rm Tr}\,\gamma_{2}^{\dagger}\gamma_{2}^{\vphantom{\dagger}}  =  {\rm Tr}\,X^{\dagger}X-2\,{\rm Det}\,X^{\dagger}X.
\label{trace4}
\end{equation}
(``Per'' denotes the permanent of a matrix.) In the derivation of Eqs. (\ref{trace2},\ref{trace3},\ref{trace4}) we have made
use of the unitarity of $S$.
The concurrence becomes
\begin{equation}
\mathcal{C}=\frac{2|\alpha|^{2}\sqrt{{\rm Det}\,X^{\dagger}X\,{\rm Det}(\openone-X^{\dagger}X)}}{
{\rm Tr}\, X^{\dagger}X-(1 + |\alpha|^{2}){\rm Per}\, X^{\dagger}X-(1 - |\alpha|^{2}){\rm Det}\, X^{\dagger}X}.
\label{Cfinal}
\end{equation}

Entanglement depends on the amount of temporal indistinguishability $|\alpha|^{2}$ and the Hermitian matrix 
\begin{equation}
X^{\dagger}X=\left( \begin{array}{cc} 
|{\mathbf r}_{\rm H}|^{2} & {\mathbf r}_{\rm H}\cdot {\mathbf t}'_{\rm V} \\ 
({\mathbf r}_{\rm H}\cdot {\mathbf t}'_{\rm V})^{*} & |{\mathbf t}'_{\rm V}|^{2} \end{array} \right),
\end{equation}
containing the states ${\mathbf r}_{\rm H}=(r_{\rm HH},r_{\rm VH})$ and ${\mathbf t}'_{\rm V}=(t'_{\rm HV},t'_{\rm VV})$ 
of a reflected and transmitted photon to the left of the beam splitter.
The determinant of $X^{\dagger}X$ measures the size of the span of ${\mathbf r}_{\rm H}$ and ${\mathbf t}'_{\rm V}$ as
\begin{equation}
{\rm Det}\, X^{\dagger}X = |{\mathbf r}_{\rm H}|^{2} |{\mathbf t}'_{\rm V}|^{2}\left(1-\frac{|{\mathbf r}_{\rm H}\cdot{\mathbf t}'_{\rm V}|^{2}}{
|{\mathbf r}_{\rm H}|^{2} |{\mathbf t}'_{\rm V}|^{2}} \right).
\end{equation}
If ${\mathbf r}_{\rm H}$ and ${\mathbf t}'_{\rm V}$ are parallel (${\rm Det}X^{\dagger}X=0$), a scattered photon to the left of the 
beam splitter is in a definite state, giving rise to an unentangled two-photon state ($\mathcal{C}=0$).
Similarly, 
\begin{equation}
{\rm Det}(\openone-X^{\dagger}X)=
|{\mathbf t}_{\rm H}|^{2} |{\mathbf r}'_{\rm V}|^{2}\left(1-\frac{|{\mathbf t}_{\rm H}\cdot {\mathbf r}'_{\rm V}|^{2}}{
|{\mathbf t}_{\rm H}|^{2} |{\mathbf r}'_{\rm V}|^{2}} \right)
\end{equation}
involves scattered states ${\mathbf t}_{\rm H}=(t_{\rm HH},t_{\rm VH})$ and ${\mathbf r}'_{\rm V}=(r'_{\rm HV},r'_{\rm VV})$ to the right
of the beam splitter. 
The denominator of Eq. (\ref{Cfinal}) is the probability of finding a scattered state with one photon on either
side of the beam splitter. It deviates from its classical value 
$(X^{\dagger}X)_{\rm HH}+(X^{\dagger}X)_{\rm VV}-2(X^{\dagger}X)_{\rm HH}(X^{\dagger}X)_{\rm VV}$
by an amount $-2|\alpha|^{2}|(X^{\dagger}X)_{\rm HV}|^{2}$ due to photon bunching. 
This reduction of coincidence count probability is the Mandel dip \cite{Hon87}. It measures the indistinguishability of a reflected and transmitted
photon as the product of temporal indistinguishability $|\alpha|^{2}$ and polarization indistinguishability $|(X^{\dagger}X)_{\rm HV}|^{2}$.     

\section{Violation of the Bell-CHSH inequality}
\label{BellCHSH}

The maximal value ${\cal E}_{\rm max}$ of the Bell-CHSH parameter (\ref{Bellpar}) for an arbitrary mixed state
was analyzed in Refs. \cite{Hor95,Ver02}. For a pure state with concurrence $\mathcal{C}$ one has simply
${\cal E}_{\rm max}=2\sqrt{1+\mathcal{C}^{2}}$ \cite{Gis91}. For a mixed state there is no one-to-one
relation between $\mathcal{C}$ and ${\cal E}_{\rm max}$. Depending on the density matrix, ${\cal E}_{\rm max}$
can take on values between $2\mathcal{C}\sqrt{2}$ and $2\sqrt{1+\mathcal{C}^{2}}$. 
The dependence of ${\cal E}_{\rm max}$ on $\rho$ involves the two largest eigenvalues 
of the real symmetric $3 \times 3$ matrix $R^{\rm T}R$ constructed from $R_{kl}={\rm Tr}\rho\, \sigma_{k}
\otimes \sigma_{l}$, where $\sigma_{1}=\sigma_{x}$,$\sigma_{2}=\sigma_{y}$ and $\sigma_{3}=\sigma_{z}$. 
In terms of $\gamma_{1}$ and $\gamma_{2}$, the elements $R_{kl}$ take the form
\begin{equation}
R_{kl}=\frac{(1+|\alpha|^{2})}{\mathcal{N}}{\rm Tr}\,\gamma_{1}^{\dagger}\sigma_{k}\gamma_{1}^{\vphantom{\dagger}}
\sigma_{l}^{\rm T}+
\frac{(1-|\alpha|^{2})}{\mathcal{N}}{\rm Tr}\,\gamma_{2}^{\dagger}\sigma_{k}\gamma_{2}^{\vphantom{\dagger}}
\sigma_{l}^{\rm T}.
\label{Rdef}
\end{equation} 

The matrix $\gamma_{2}$ has a polar decomposition $\gamma_{2}=U\sqrt{\xi}V$ where $U$ and $V$ are unitary
matrices and $\xi$ is a diagonal matrix holding the eigenvalues of 
$\gamma_{2}^{\dagger}\gamma_{2}^{\vphantom{\dagger}}$. 
The real positive $\xi_{i}$'s are determined by
\begin{equation}
\xi_{1}+\xi_{2}={\rm Tr}\,\gamma_{2}^{\dagger}\gamma_{2}^{\vphantom{\dagger}}, \quad 
2\sqrt{\xi_{1}\xi_{2}}=|{\rm Tr}\,\gamma_{2}^{\dagger}\tilde{\gamma}_{2}^{\vphantom{\dagger}}|.
\label{xi}
\end{equation}
The matrix $\gamma_{1}$ can be conveniently expressed as (see Appendix \ref{semipol})
\begin{equation}
\gamma_{1}=UQ\sqrt{\xi}V, \quad \mbox{where} \quad Q=\left( \begin{array}{cc} c_{1} & c_{2} \\
c_{3} & -c_{1} \end{array}\right). 
\label{gam1}
\end{equation}
The parameters $c_{1}$,$c_{2}$,$c_{3}$ are real numbers. The matrix $Q$ is traceless due to the orthogonality of $\gamma_{1}$ and
$\tilde{\gamma}_{2}$. The number $c_{1} \in (-1,1)$ on the diagonal is related to the inner product of
$\gamma_{1}$ and $\gamma_{2}$ and takes the form
\begin{equation}
c_{1}=\frac{{\rm Tr}\,\gamma_{1}^{\dagger}\gamma_{2}^{\vphantom{\dagger}}}{\xi_{1}-\xi_{2}}, \quad \mbox{with} \quad 
{\rm Tr}\,\gamma_{1}^{\dagger}\gamma_{2}^{\vphantom{\dagger}}={\rm Tr}\,\sigma_{z}X^{\dagger}X.
\label{c1}
\end{equation}
The numbers $c_{2}$,$c_{3}$ are determined by the norm
and tilde inner product of $\gamma_{1}$ and satisfy the relations 
\begin{equation}
c_{1}^{2}+c_{2}c_{3}=1, \quad c_{1}^{2}(\xi_{1}+\xi_{2})+c_{2}^{2}\xi_{2}+c_{3}^{2}\xi_{1}=
{\rm Tr}\, \gamma_{1}^{\dagger}\gamma_{1}^{\vphantom{\dagger}}.
\label{c2&c3}
\end{equation}

We substitute $\gamma_{1}$ of Eq. (\ref{gam1}) and the polar decomposition of $\gamma_{2}$ in Eq. (\ref{Rdef}) and parameterize
\begin{equation}
U^{\dagger}\sigma_{k} U = \sum_{i=1}^{3} N_{ki}\sigma_{i}, \quad 
V\sigma_{k}^{\rm T} V^{\dagger} = \sum_{i=1}^{3} M_{ki} \sigma_{i}^{\rm T},
\end{equation} 
in terms of two $3 \times 3$ orthogonal matrices $N$ and $M$.   
The matrix $R$ takes the form
\begin{equation}
R=N R' M^{\rm T},
\end{equation}
where $R'$ is given by Eq. (\ref{Rdef}) with substitutions $R \rightarrow R'$, 
$\gamma_{2} \rightarrow \sqrt{\xi}$ and $\gamma_{1} \rightarrow Q\sqrt{\xi}$.
With the help of Eqs. (\ref{xi},\ref{c1},\ref{c2&c3}), the eigenvalues $u_{i}$ of 
$R^{\rm T}R$ can now be expressed as (see Appendix \ref{RtReig})
\begin{equation}
u_{1}=\frac{1}{2\mathcal{N}^{2}}\left(\mathcal{T}+\sqrt{\mathcal{T}^{2}-4\mathcal{D}}\right),
\label{u1}
\end{equation}
\begin{equation}
u_{2}=\frac{1}{2\mathcal{N}^{2}}\left(\mathcal{T}-\sqrt{\mathcal{T}^{2}-4\mathcal{D}}\right),
\label{u2}
\end{equation}
\begin{equation}
u_{3}=4\frac{|\alpha|^{4}|{\rm Tr}\,\gamma_{1}^{\dagger}\tilde{\gamma}_{1}^{\vphantom{\dagger}}|^{2}}{\mathcal{N}^{2}},
\label{u3}
\end{equation}
where 
\begin{widetext}
\begin{equation}
\mathcal{T}=\mathcal{N}^{2}+4|{\rm Tr}\,\gamma_{1}^{\dagger}\tilde{\gamma}_{1}^{\vphantom{\dagger}}|^{2}
-4(1-|\alpha|^{4})\left({\rm Tr}\,\gamma_{1}^{\dagger}\gamma_{1}^{\vphantom{\dagger}}
{\rm Tr}\,\gamma_{2}^{\dagger}\gamma_{2}^{\vphantom{\dagger}}-
{\rm Tr}^{2}\,\gamma_{1}^{\dagger}\gamma_{2}^{\vphantom{\dagger}}\right),
\label{mathT}
\end{equation}
\begin{equation}
\mathcal{D}=4|{\rm Tr}\,\gamma_{1}^{\dagger}\tilde{\gamma}_{1}^{\vphantom{\dagger}}|^{2}\left(\mathcal{N}^{2}-4(1-|\alpha|^{4})
{\rm Tr}\,\gamma_{1}^{\dagger}\gamma_{1}^{\vphantom{\dagger}}
{\rm Tr}\,\gamma_{2}^{\dagger}\gamma_{2}^{\vphantom{\dagger}}\right).
\label{mathD}
\end{equation}
\end{widetext}
We can relate the $u_{i}$'s to $X^{\dagger}X$ and $|\alpha|^{2}$ using Eqs.
(\ref{norm},\ref{trace2},\ref{trace3},\ref{trace4},\ref{c1}). 
The parameter ${\cal E}_{\rm max}$ depends on the two largest eigenvalues of $R^{\rm T}R$ as
\begin{equation}
{\cal E}_{\rm max}=2\sqrt{u_{1}+{\rm max}(u_{2},u_{3})}.
\label{Emax}
\end{equation} 
Generically, the expression for ${\cal E}_{\rm max}$ takes a complicated form where ordering of $u_{2}$ and
$u_{3}$ depends on $X^{\dagger}X$ and $|\alpha|^{2}$.

\section{Discussion}
\label{Dis}

The objective of the discussion is to reveal the role played by the Mandel dip $-2|\alpha|^{2}|(X^{\dagger}X)_{\rm HV}|^{2}$
in the connection between $\mathcal{C}$ and ${\cal E}_{\rm max}$.

We first consider the case $|(X^{\dagger}X)_{\rm HV}|^{2}=0$. The concurrence of Eq. (\ref{Cfinal}) reduces to     
\begin{equation}
\mathcal{C}=\frac{2|\alpha|^{2}\prod_{i={\rm H,V}}\sqrt{(X^{\dagger}X)_{ii}(1-(X^{\dagger}X)_{ii})}}{(X^{\dagger}X)_{\rm HH}+(X^{\dagger}X)_{\rm VV}
-2(X^{\dagger}X)_{\rm HH}(X^{\dagger}X)_{\rm VV}}.
\end{equation}
The maximal value of the Bell-CHSH parameter takes the form   
\begin{equation}
{\cal E}_{\rm max}=2\sqrt{1+\mathcal{C}^{2}}
\label{Enomix}
\end{equation}
and $\mathcal{C} > 0$ implies ${\cal E}_{\rm max} > 2$.

\begin{figure}
\includegraphics[width=8cm]{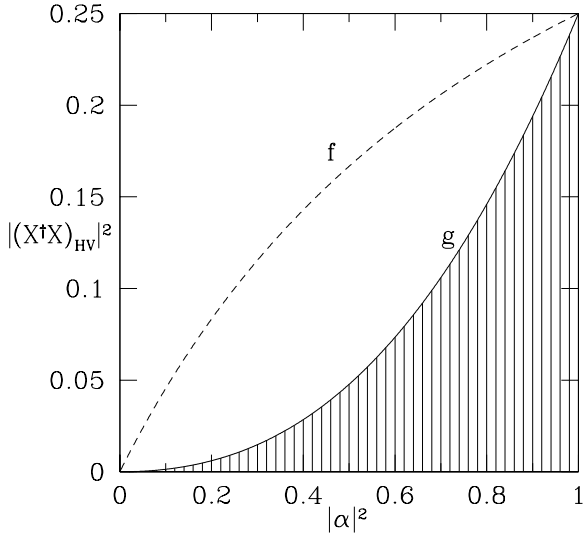}
\caption{
Parameter space of a beam splitter with $(X^{\dagger}X)_{ii}=1/2$ spanned by $|\alpha|^{2} \in (0,1)$ and $|(X^{\dagger}X)_{\rm HV}|^{2} \in (0,1/4)$. 
All points correspond to a non-vanishing polarization-entanglement ($\mathcal{C} > 0$) except the line segments $|\alpha|^{2}=0$ and
$|(X^{\dagger}X)_{\rm HV}|^{2}=1/4$ where entanglement vanishes ($\mathcal{C}=0$). Only in the shaded region, the Bell-CHSH parameter is
able to detect entanglement (${\cal E}_{\rm max} > 2$). The lines correspond to the functions $f$, $g$ of Eqs. (\ref{f},\ref{g})
respectively.   
\label{regions}
}
\end{figure}

In the presence of a Mandel dip ($|\alpha|^{2}|(X^{\dagger}X)_{\rm HV}|^{2} > 0$), the ability of ${\cal E}$ to witness entanglement can
disappear. We consider the special case $(X^{\dagger}X)_{ii}=1/2$.
The concurrence of Eq. (\ref{Cfinal}) reduces to 
\begin{equation}
\mathcal{C}=\frac{|\alpha|^{2}\left(1-4|(X^{\dagger}X)_{\rm HV}|^{2}\right)}{1-4|\alpha|^{2}|(X^{\dagger}X)_{\rm HV}|^{2}}. 
\end{equation}
To find ${\cal E}_{\rm max}$ we have to consider the ordering of $u_{2}$ and $u_{3}$ which depends on $|(X^{\dagger}X)_{\rm HV}|^{2}$ and $|\alpha|^{2}$.
The function 
\begin{equation}
f(|\alpha|^{2})=\frac{|\alpha|^{2}}{2(1+|\alpha|^{2})}
\label{f}
\end{equation}
divides parameter space in the region $|(X^{\dagger}X)_{\rm HV}|^{2} \le f$ where 
${\cal E}_{\rm max}=2\sqrt{u_{1}+u_{3}}$ and the region
$|(X^{\dagger}X)_{\rm HV}|^{2} > f$ where ${\cal E}_{\rm max}=2\sqrt{u_{1}+u_{2}}$. The equation ${\cal E}_{\rm max}=2$ has a solution
$g(|\alpha|^{2})$ for $|(X^{\dagger}X)_{\rm HV}|^{2}$ that lies in the region $|(X^{\dagger}X)_{\rm HV}|^{2} \le f$. The function $g$ takes the form
\begin{equation}
g(|\alpha|^{2})=\frac{1}{4}\left(1-|\alpha|^{2}+|\alpha|^{4}-(1-|\alpha|^{2})\sqrt{1+|\alpha|^{4}}\right)
\label{g}
\end{equation}
and breaks parameter space in two fundamental regions: a region $|(X^{\dagger}X)_{\rm HV}|^{2}<g$ where ${\cal E}_{\rm max} > 2$
and a region $|(X^{\dagger}X)_{\rm HV}|^{2}>g$ where ${\cal E}_{\rm max} < 2$.
We have drawn these regions in Fig. \ref{regions}. 
The maximal value of the Bell-CHSH parameter is given by
\begin{equation}
{\cal E}_{\rm max}=2\mathcal{C}|\alpha|^{-2}\sqrt{1+|\alpha|^{4}} 
\end{equation}  
in the region $|(X^{\dagger}X)_{\rm HV}|^{2} \le f$. 

\section{Conclusions}

In summary, we have calculated the amount of polarization-entanglement (concurrence $\mathcal{C}$) and its witness 
(maximal value of the Bell-CHSH parameter ${\cal E}$) induced by two-photon interference at a lossless beam splitter.
The ability of ${\cal E}$ to witness entanglement (${\cal E}_{\rm max} > 2$) depends on the Mandel dip $-2|\alpha|^{2}|(X^{\dagger}X)_{\rm HV}|^{2}$.
In the absence of a Mandel dip, $\mathcal{C} > 0$ implies ${\cal E}_{\rm max} > 2$ cf. Eq. (\ref{Enomix}), whereas in its presence this is not
necessarily true. In the latter case, as we have demonstrated in Sec. \ref{Dis} with $(X^{\dagger}X)_{ii}=1/2$, the witnessing ability of ${\cal E}$
depends on the individual contributions of temporal ($|\alpha|^{2}$) and polarization indistinguishability 
($|(X^{\dagger}X)_{\rm HV}|^{2}$). 

Our results can be applied to interference of other kinds of particles, getting entangled in some $2 \otimes 2$ Hilbert space and being ``marked''
by an additional degree of freedom. However, determining the indistinguishability parameter $|\alpha|^{2}$ requires careful analysis of the detection scheme.
In case of fermions, the matrices $\gamma_{1}$ and $\gamma_{2}$ of Eq. (\ref{gammas}) are to be interchanged. 
Systems without a time-reversal symmetry are captured by
the analysis, as we did not make use of the symmetry of the scattering matrix.     

\acknowledgments
I am grateful to C. W. J. Beenakker for discussions and advice.
This work was supported by the Dutch Science Foundation NWO/FOM and by the U.S. Army Research Office (Grant No. DAAD 19-02-0086).

\appendix

\section{arbitrariness of two-photon input state}
\label{stostate}

The unitary scattering matrix has a polar decomposition
\begin{equation}
S=\left(\begin{array}{cc} K' & 0 \\ 0 & L' \end{array}\right)
\left(\begin{array}{cc} \sqrt{\openone-T} & i\sqrt{T} \\ i\sqrt{T} & \sqrt{\openone-T} \end{array}\right)
\left(\begin{array}{cc} K & 0 \\ 0 & L \end{array}\right),
\label{Spolar}
\end{equation}
where $K'$,$L'$,$K$,$L$ are $2 \times 2$ unitary matrices and $T={\rm diag}(T_{\rm H},T_{\rm V})$ is a matrix of transmission eigenvalues 
$T_{\rm H},T_{\rm V} \in (0,1)$.
The outgoing state $|\Psi_{\rm out}\rangle$ is related to the $4 \times 4$ matrix 
\begin{equation}
S \left( \begin{array}{cc} 0 & \sigma_{\rm in} \\ 0 & 0 \end{array} \right) S^{\rm T}
\end{equation}
cf. Eq. (\ref{psimatout}).
By group decomposition $K=K_{1}K_{2}$ and $L=L_{1}L_{2}$, $|\Psi_{\rm out}\rangle$ is 
easily seen to correspond to $K_{2}\sigma_{\rm in}L_{2}^{\rm T}$ scattered by $S$ of Eq. (\ref{Spolar}) with substitutions 
$K \rightarrow K_{1}$ and $L \rightarrow L_{1}$.  

\section{joint semi-polar decomposition}
\label{semipol}

The matrices $\gamma_{1}$ and $\gamma_{2}$ have a decomposition
\begin{equation}
\gamma_{1}=U\mathcal{A}V, \quad \gamma_{2}=U\sqrt{\xi}V,
\label{jointdec}
\end{equation}
where $U$,$V$ are unitary matrices and $\xi$ is a diagonal matrix of eigenvalues 
of $\gamma_{2}^{\dagger}\gamma_{2}^{\vphantom{\dagger}}$. 
As we do not yet specify $\mathcal{A}$, such a joint decomposition always exists.
In our case, the matrices $\gamma_{1}$ and $\gamma_{2}$ have the special properties
\begin{equation}
{\rm Tr}\,\gamma_{1}^{\dagger}\tilde{\gamma}_{1}^{\vphantom{\dagger}}=-
{\rm Tr}\,\gamma_{2}^{\dagger}\tilde{\gamma}_{2}^{\vphantom{\dagger}}, \quad
{\rm Tr}\,\gamma_{1}^{\dagger}\tilde{\gamma}_{2}^{\vphantom{\dagger}}=0, \quad
\end{equation}
\begin{equation}
|{\rm Tr}\,\gamma_{1}^{\dagger}\tilde{\gamma}_{1}^{\vphantom{\dagger}}|  = 
2\sqrt{{\rm Det}\,X^{\dagger}X\,{\rm Det}(\openone-X^{\dagger}X)}, 
\end{equation}
\begin{equation}
{\rm Tr}\,\gamma_{1}^{\dagger}\gamma_{1}^{\vphantom{\dagger}}  =  {\rm Tr}\,X^{\dagger}X-2\,{\rm Per}\,X^{\dagger}X, 
\end{equation}
\begin{equation}
{\rm Tr}\,\gamma_{2}^{\dagger}\gamma_{2}^{\vphantom{\dagger}}  =  {\rm Tr}\,X^{\dagger}X-2\,{\rm Det}\,X^{\dagger}X,
\end{equation}
\begin{equation}
{\rm Tr}\,\gamma_{1}^{\dagger}\gamma_{2}^{\vphantom{\dagger}}={\rm Tr}\,\sigma_{z}X^{\dagger}X.
\end{equation}
It is the purpose of this appendix to demonstrate that $\mathcal{A}=Q\sqrt{\xi}$ where
$Q$ is a real traceless matrix of Eq. (\ref{gam1}) with $c_{1}$ given by Eq. (\ref{c1}) and $c_{2}$,$c_{3}$ 
satisfying Eq. (\ref{c2&c3}). 

The inner and tilde inner product of $\gamma_{1}$ and $\gamma_{2}$ take the form
\begin{equation}
{\rm Tr}\,\gamma_{1}^{\dagger}\gamma_{2}^{\vphantom{\dagger}}={\rm Tr}\,\mathcal{A}^{\dagger}\sqrt{\xi}, 
\label{inner1}
\end{equation}
\begin{equation}
{\rm Tr}\,\gamma_{1}^{\dagger}\tilde{\gamma}_{2}^{\vphantom{\dagger}}=({\rm Det}\, UV)^{-2}\,{\rm Tr}\,\mathcal{A}^{\dagger}\sigma_{y}
\sqrt{\xi}\sigma_{y}=0.
\label{inner2}
\end{equation}
(Here we have used the identity $U\sigma_{y}U^{\rm T}={\rm Det}^{2}U\, \sigma_{y}$, valid for any $2 \times 2$ unitary matrix $U$.)
The conditions of Eqs. (\ref{inner1},\ref{inner2}) involve the diagonal elements of $\mathcal{A}$ as respectively
\begin{equation}
{\rm Tr}\,\gamma_{1}^{\dagger}\gamma_{2}^{\vphantom{\dagger}}=\sqrt{\xi_{1}}\mathcal{A}_{11}^{*}+\sqrt{\xi_{2}}\mathcal{A}_{22}^{*},
\end{equation}
\begin{equation}
{\rm Tr}\,\gamma_{1}^{\dagger}\tilde{\gamma}_{2}^{\vphantom{\dagger}}=
({\rm Det}\, UV)^{-2}
\left(\sqrt{\xi_{1}}\mathcal{A}_{22}^{*}+\sqrt{\xi_{2}}\mathcal{A}_{11}^{*}\right)=0.
\end{equation}
It follows that $\mathcal{A}_{11}=c_{1}^{*}\sqrt{\xi_{1}}$ and $\mathcal{A}_{22}=-c_{1}^{*}\sqrt{\xi_{2}}$ where $c_{1}$ is given by 
\begin{equation}
c_{1}=\frac{{\rm Tr}\,\gamma_{1}^{\dagger}\gamma_{2}^{\vphantom{\dagger}}}{\xi_{1}-\xi_{2}}.
\end{equation}
The number $c_{1}$ is real since
${\rm Tr}\,\gamma_{1}^{\dagger}\gamma_{2}^{\vphantom{\dagger}}={\rm Tr}\,\sigma_{z}X^{\dagger}X \in \mathbb{R}$.

The determinant of $\mathcal{A}$ is fixed by ${\rm Tr}\,\gamma_{1}^{\dagger}\tilde{\gamma}_{1}^{\vphantom{\dagger}}=-
{\rm Tr}\,\gamma_{2}^{\dagger}\tilde{\gamma}_{2}^{\vphantom{\dagger}}$ implying
\begin{equation}
{\rm Det}\,\mathcal{A}=-\sqrt{\xi_{1}\xi_{2}}.
\end{equation}
It follows that $\mathcal{A}_{12}=\mathcal{A}'_{12}e^{i\phi}$ and $\mathcal{A}_{21}=\mathcal{A}'_{21}e^{-i\phi}$ with real 
$\mathcal{A}'_{12}$,$\mathcal{A}'_{21}$,$\phi$. The numbers $\mathcal{A}'_{12}$,$\mathcal{A}'_{21}$ satisfy 
\begin{equation}
c_{1}^{2}\sqrt{\xi_{1}\xi_{2}}+\mathcal{A}'_{12}\mathcal{A}'_{21}=\sqrt{\xi_{1}\xi_{2}},
\label{detA}
\end{equation}
\begin{equation}
c_{1}^{2}(\xi_{1}+\xi_{2})+{\mathcal{A}'}_{12}^{2}+{\mathcal{A}'}_{21}^{2}=
{\rm Tr}\,\gamma_{1}^{\dagger}\gamma_{1}^{\vphantom{\dagger}},
\label{TrA}
\end{equation}
where Eq. (\ref{TrA}) comes from ${\rm Tr}\,\mathcal{A}^{\dagger}A={\rm Tr}\,\gamma_{1}^{\dagger}\gamma_{1}^{\vphantom{\dagger}}$.
The undetermined phase $\phi$ can be taken out,
\begin{equation}
\mathcal{A} = \left(\begin{array}{cc} e^{i\frac{\phi}{2}} & 0 \\ 0 & e^{-i\frac{\phi}{2}} \end{array} \right)
\left(\begin{array}{cc} \mathcal{A}_{11} & \mathcal{A}'_{12} \\ \mathcal{A}'_{21} & \mathcal{A}_{22} \end{array} \right)
\left(\begin{array}{cc} e^{-i\frac{\phi}{2}} & 0 \\ 0 & e^{i\frac{\phi}{2}} \end{array} \right),
\end{equation}  
and absorbed in the unitary matrices $U$ and $V$ by the transformations
\begin{equation}
U\left(\begin{array}{cc} e^{i\frac{\phi}{2}} & 0 \\ 0 & e^{-i\frac{\phi}{2}} \end{array} \right) \rightarrow U, \quad
\left(\begin{array}{cc} e^{-i\frac{\phi}{2}} & 0 \\ 0 & e^{i\frac{\phi}{2}} \end{array} \right)V \rightarrow V.
\end{equation}
(Note that these transformations also hold for $\gamma_{2}$ since $\sqrt{\xi}$ commutes with 
a diagonal matrix of phase factors.)

The matrix $\mathcal{A}$ is related to $Q$ by $\mathcal{A}=Q\sqrt{\xi}$.
It is now easily seen that the matrix $Q$ is real and traceless and takes the form of Eq. (\ref{gam1}), with $c_{1}$ 
given by Eq. (\ref{c1}) and $c_{2}$,$c_{3}$ satisfying Eq. (\ref{c2&c3}).

As a last step we perform a consistency check to demonstrate that Eqs. (\ref{detA},\ref{TrA}) have solutions for 
$\mathcal{A}'_{12}$ and $\mathcal{A}'_{21}$. 
The Hermitian matrix $X^{\dagger}X$ has an eigenvalue-eigenvector decomposition
\begin{equation}
X^{\dagger}X=W^{\dagger}\Lambda W.
\end{equation}
In terms of the eigenvalues $\Lambda_{i} \in (0,1)$ and the unitary matrix $W$,
the inner product of $\gamma_{1}$ and $\gamma_{2}$ and the $\xi_{i}$'s take the form
\begin{equation}
{\rm Tr}\,\gamma_{1}^{\dagger}\gamma_{2}^{\vphantom{\dagger}}=\Lambda_{1}(|W_{11}|^{2}-|W_{12}|^{2})
+\Lambda_{2}(|W_{21}|^{2}-|W_{22}|^{2}), 
\end{equation}
\begin{equation}
\xi_{1}=\Lambda_{1}(1-\Lambda_{2}), \quad \xi_{2}=\Lambda_{2}(1-\Lambda_{1}).
\end{equation}
It follows that $c_{1}=\cos 2\eta$, where we have set $|W_{11}|=|W_{22}|=\cos \eta$ and $|W_{12}|=|W_{21}|=\sin \eta$.
Eqs. (\ref{detA},\ref{TrA}) can be expressed as respectively
\begin{equation}
\mathcal{A}'_{12}\mathcal{A}'_{21}=\sin^{2}2\eta\, \sqrt{\Lambda_{1}\Lambda_{2}(1-\Lambda_{1})(1-\Lambda_{2})},
\end{equation}
\begin{equation}
{\mathcal{A}'}_{12}^{2}+{\mathcal{A}'}_{21}^{2}=\sin^{2}2\eta\, (\Lambda_{1}(1-\Lambda_{1})+\Lambda_{2}(1-\Lambda_{2})).
\end{equation}
Since 
\begin{equation}
2\sqrt{\Lambda_{1}\Lambda_{2}(1-\Lambda_{1})(1-\Lambda_{2})} \le \Lambda_{1}(1-\Lambda_{1})+\Lambda_{2}(1-\Lambda_{2})
\end{equation}
a family of solutions exists.

\section{eigenvalues of ${\mathbf R}^{\rm {\mathbf T}}{\mathbf R}$}
\label{RtReig}

The non-vanishing elements of $R'$ are given by
\begin{equation}
R'_{11}=\frac{2}{\mathcal{N}}
\left(1-|\alpha|^{2}-(1+|\alpha|^{2})(c_{1}^{2}-c_{2}c_{3})\right)\sqrt{\xi_{1}\xi_{2}}, 
\end{equation}
\begin{equation}
R'_{13}=\frac{2}{\mathcal{N}}(1+|\alpha|^{2})c_{1}(c_{2}\xi_{2}+c_{3}\xi_{1}),
\end{equation}
\begin{equation}
R'_{22}=\frac{2}{\mathcal{N}}\left(-1+|\alpha|^{2}+(1+|\alpha|^{2})(c_{1}^{2}+c_{2}c_{3})\right)\sqrt{\xi_{1}\xi_{2}},
\end{equation}
\begin{equation}
R'_{31}=\frac{2}{\mathcal{N}}(1+|\alpha|^{2})c_{1}(c_{2}+c_{3})\sqrt{\xi_{1}\xi_{2}},
\end{equation}
\begin{eqnarray}
R'_{33}=\frac{1}{\mathcal{N}}\left((1-|\alpha|^{2})+(1+|\alpha|^{2})c_{1}^{2}\right)(\xi_{1}+\xi_{2}) \\
-\frac{1}{\mathcal{N}}(1+|\alpha|^{2})(c_{2}^{2}\xi_{2}+c_{3}^{2}\xi_{1}).
\end{eqnarray}
The matrix $R'^{\rm T}R'$ has eigenvalues 
\begin{equation}
u_{1}=\frac{1}{2\mathcal{N}^{2}}\left(\mathcal{T}+\sqrt{\mathcal{T}^{2}-4\mathcal{D}}\right),
\end{equation}
\begin{equation}
u_{2}=\frac{1}{2\mathcal{N}^{2}}\left(\mathcal{T}-\sqrt{\mathcal{T}^{2}-4\mathcal{D}}\right),
\end{equation}
\begin{equation}
u_{3}={R'}_{22}^{2}, 
\end{equation}
where $\mathcal{T}$,$\mathcal{D}$ are the trace, determinant respectively of the $2 \times 2$ real symmetric matrix
\begin{equation}
\mathcal{N}^{2}\left(\begin{array}{cc} {R'}_{11}^{2}+{R'}_{31}^{2} & {R'}_{11}{R'}_{13}+{R'}_{31}{R'}_{33} \\
{R'}_{11}{R'}_{13}+{R'}_{31}{R'}_{33} & {R'}_{13}^{2}+{R'}_{33}^{2} \end{array}\right).
\end{equation}  
By making use of Eqs. (\ref{c1},\ref{c2&c3}) $u_{3}$,$\mathcal{T}$,$\mathcal{D}$ can be simplified to yield the results of
Eqs. (\ref{u3},\ref{mathT},\ref{mathD}) respectively.

\end{document}